\documentclass[aps,prl,showpacs,twocolumn]{revtex4}

\usepackage{epsfig,amsmath,color,amssymb}

\begin{document}

\title{A Deterministic and Storable Single-Photon Source Based on Quantum Memory}

\author{Shuai Chen}
\thanks{These authors contribute equally to this work.}
\author{Yu-Ao Chen}
\thanks{These authors contribute equally to this work.}
\author{Thorsten Strassel}
\thanks{These authors contribute equally to this work.}
\author{Zhen-Sheng Yuan}
\thanks{These authors contribute equally to this work.}
\author{Bo Zhao}
\author{J\"{o}rg Schmiedmayer}
\author{Jian-Wei Pan}

\address{Physikalisches Institut, Ruprecht-Karls-Universit\"{a}t Heidelberg, Philosophenweg 12,
69120 Heidelberg, Germany}
\date{\today}

\begin{abstract}

A single photon source is realized with a cold atomic ensemble ($^{87}$Rb atoms). In the
experiment, single photons, which is initially stored in an atomic quantum memory generated by
Raman scattering of a laser pulse, can be emitted deterministically at a time-delay in control. It
is shown that production rate of single photons can be enhanced by a feedback circuit considerably
while the single-photon quality is conserved. Thus our present single-photon source is well
suitable for future large-scale realization of quantum communication and linear optical quantum
computation.

\end{abstract}



\maketitle

Although weak coherent beams can be used as a pseudo single-photon source, the advent of quantum
information processing (QIP) has placed stringent requirements on single photons either on demand
or heralded \cite{LounisRRP2005}. In particular, secure quantum cryptography \cite{GisinRMP2002}
and linear optical quantum computing \cite{KnillNature2001} depend on the availability of such
single-photon sources. Different approaches have been attempted in the last decade to develop the
on-demand single-photon source, such as the implementations with quantum dots \cite{MichlerSCI2000,
SantoriPRL2001}, single atoms and ions \cite{KuhnPRL2002, KellerNature2004}, and color centers
\cite{KurtsieferPRL2000}. However, all of them are confronted with different challenges. For
example, the single-atom implementation provides spectrally narrow single photons with a well
defined spatial mode, but the main challenge is the manipulation of single atoms, which requires
sophisticated and expensive setups \cite{KuhnPRL2002}. Although quantum dots present many
advantages as potential source of single photons, e.g. high single-photon rate, the requirement of
spectral filtering entails inevitable losses. Additionally, it is a major problem for preparing
truly identical sources due to inhomogeneities in both the environment of the emitters and the
emitters itself \cite{SantoriNJP2004}. The stability of color centers is excellent, even at room
temperature. However, the high peak intensities of a pulsed excitation can lead to complex and
uncontrollable dark states \cite{LounisRRP2005}. So it has been taken as a formidable task to
develop a promising single-photon source.

Moreover, an important challenge in distributed QIP is the controllable transfer of quantum state
between flying qubit and macroscopic matter. Remarkably, as shown in a recent proposal for
long-distance quantum communication with atomic ensembles \cite{DuanNature2001}, it is possible to
implement both a single-photon source on demand and the controllable transfer of quantum state
between photonic qubit and macroscopic matter, provided that proper feedback is applied to achieve
the classical feed-forward ability. Such feed-forward ability is a crucial requirement in linear
optics QIP \cite{KnillNature2001,DuanNature2001}. In other words, it must be, in principle,
possible to detect when an operation has succeeded by performing some appropriate measurement on
ancilla photons. This information can then be feed-forwarded for conditional future operations on
the photonic qubits to achieve efficient QIP.

Recently, significant experimental progresses have been achieved in demonstration of quantum
storage and single-photon sources \cite{KuzmichNature2003, ChouPRL2004, ChaneliereNature2005,
EisamanNature2005}, and even entanglement between two atomic ensembles \cite{MatsukevichSCI2004,
ChouNature2005} has been generated. However, coincidence-based post-selection was used in these
experiments. Consequently no feedback could be applied to achieve the feed-forward ability and the
requirement of resources will exponentially increase with each new step of operation. This
significantly limit the scalability of the schemes \cite{KnillNature2001,DuanNature2001}.

In this letter, we present an experimental realization of a deterministic and storable
single-photon source. Single spin excitations in an atomic ensemble are generated by detecting
anti-Stokes photons from spontaneous Raman scattering. This detection allows to implement
feed-forward and convert the spin excitations into single photons at a predetermined given time.
Moreover, it is shown that the single-photon quality is conserved while the production rate of
single photons can be enhanced greatly by the feedback circuit. In principle, the spatial mode,
bandwidth, and frequency of single-photon pulses are determined by the spatial mode, intensity and
frequency of the retrieve laser \cite{EisamanNature2005}. So it is feasible to integrate such a
single-photon source with the storage medium, atomic ensembles. Together with the technology
developed in previous experiments \cite{KuzmichNature2003, ChouPRL2004, ChaneliereNature2005,
EisamanNature2005, MatsukevichSCI2004, ChouNature2005}, our controllable single-photon source
potentially paves the way for the construction of scalable quantum communication networks
\cite{DuanNature2001, BriegelPRL1998} and linear optical quantum computation
\cite{KnillNature2001}.

\begin{figure}[tb]
\includegraphics[width=8.0cm]{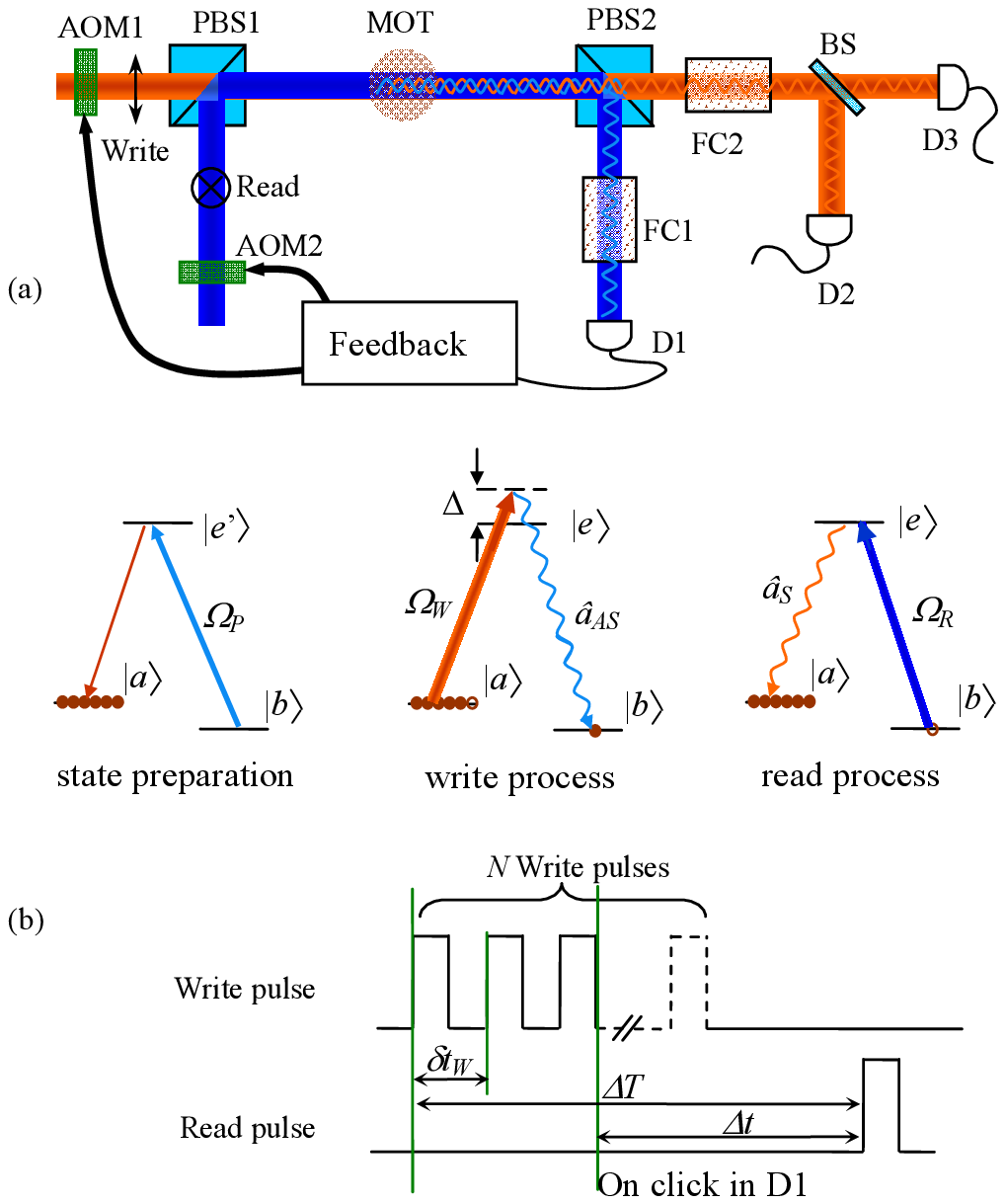}
\caption{Illustration of the experimental setup (a) and the time sequence with the feedback circuit
for the \textit{write} and \textit{read} process (b). (a) The cold $^{87}$Rb atoms is prepared in
the MOT with the state $|a\rangle=|5S_{1/2},F=2\rangle$ (the other three levels $|b\rangle$,
$|e\rangle$, $|e'\rangle$ correspond to $|5S_{1/2},F=1\rangle$, $|5P_{1/2},F=2\rangle$ and
$|5P_{3/2},F=2\rangle$, respectively). The atomic ensemble is firstly prepared in the initial state
$|a\rangle$ by applying a pump beam resonant with the transition $|b\rangle$ to $|e'\rangle$. A
write pulse with the Rabi frequency $\Omega_{\scriptscriptstyle W}$ is applied to generate the spin
excitation and an accompanying photon of the mode $\hat{a}_{\scriptscriptstyle AS}$. Waiting for a
duration $\delta t$, a read pulse is applied with orthogonal polarization and spatially overlap
with the write beam in PBS1. The photons, whose polarization is orthogonal to that of the write
beam, in the mode $\hat{a}_{\scriptscriptstyle AS}$ are spatially extracted from the write beam by
PBS2 and detected by detector D1. Similarly, the field $\hat{a}_{\scriptscriptstyle S}$ is
spatially extracted from the Read beam and detected by detector D2 (or D3). Here, FC1 and FC2 are
two filter cells filled with $^{87}$Rb atoms to filter the leakage of write beam and read beam
respectively. (b) The time sequence can be achieved with the feedback circuit and the two
acousto-optic modulators AOM1 and AOM2.\label{fig:setup}}
\end{figure}

The basic concept of our experiments is shown in Fig. \ref{fig:setup}. Cold atoms with
$\Lambda$-type level configuration (two ground state $|a\rangle$, $|b\rangle$ and an excited state
$|e\rangle$) collected by a magneto-optical trap (MOT) are used as the media for quantum memory.
The atoms are initially optically pumped to state $|a\rangle$ by a pump laser. Then a weak
classical \textit{write} pulse, with the Rabi frequency $\Omega_{\scriptscriptstyle W}$, close to
the resonance of transition $|a\rangle$ to $|e\rangle$ is introduced in the atomic cloud. Due to
the spontaneous Raman process, a photon of anti-Stokes field $\hat{a}_{\scriptscriptstyle AS}$ is
emitted into the forward scattering mode. Simultaneously, a collective spin excitation
corresponding to the mode of the anti-Stokes field $\hat{a}_{\scriptscriptstyle AS}$ is generated
in the atomic ensemble \cite{DuanNature2001, LukinRMP2003}. The state of the field
$\hat{a}_{\scriptscriptstyle AS}$ and the collective spin state of the atoms can be expressed by
the superposed state
\begin{eqnarray}\label{eqn:state}
|\Psi\rangle=|0_{\scriptscriptstyle AS}0_b\rangle+ \sqrt{\chi}|1_{\scriptscriptstyle
 AS}1_b\rangle+ \chi|2_{\scriptscriptstyle  AS}2_b\rangle+O(\chi^{3/2}),
\end{eqnarray}
where $\chi$ is the excitation probability of one spin flip, $|i_{\scriptscriptstyle AS}i_b\rangle$
denotes the $i$-fold excitation of the anti-Stokes field and the collective spin. Ideally,
conditioned on detecting one and only one anti-Stokes photon in detector D1, a single spin
excitation is generated in the atomic ensemble with certainty. After a controllable time delay
$\delta t$ (in the order of the lifetime $\tau_c$ of the spin excitation), another classical
\textit{read} pulse with the Rabi frequency $\Omega_{\scriptscriptstyle R}$, which is on-resonance
with the transition from $|b\rangle$ to $|e\rangle$, is applied to retrieve the spin excitation and
generate a photon of Stokes field $\hat{a}_{\scriptscriptstyle S}$.

In our present experiment, more than $10^8$ $^{87}$Rb atoms are collected by the MOT with an
optical depth of about 5 and the temperature of about 100 $\mu$K. The earth magnetic field is
compensated by three pairs of Helmholtz coils. The two ground states $|a\rangle$ and $|b\rangle$
and the excited state $|e\rangle$ in the $\Lambda$-type system are
$|5S_{1/2},F=2\rangle,~|5S_{1/2},F=1\rangle$, and $|5P_{1/2},F=2\rangle$, respectively. The write
laser is tuned to the transition from $|5S_{1/2},F=2\rangle$ to $|5P_{1/2},F=2\rangle$ with
detuning of $10$ MHz and the read laser is locked on resonance to the transition from
$|5S_{1/2},F=1\rangle$ to $|5P_{1/2},F=2\rangle$. By using orthogonal polarizations, write and read
beams are spatially overlapped on a polarized beam splitter (PBS1), and then focused into the cold
atoms with the beam waist of 35 $\mu$m. After going through the atomic cloud, the two beams are
split by PBS2 which also serves as the first stage of filtering the write (read) beam out from the
anti-Stokes (Stokes) field. The leakage of write (read) field from PBS2 propagating with the
anti-Stokes (Stokes) field will be further filtered by a thermal cell filled with $^{87}$Rb atoms,
in which the rubidium atoms are prepared in state $|5S_{1/2},F=2\rangle$ ($|5S_{1/2},F=1\rangle$)
initially. Coincident measurements among D1, D2 and D3 are performed with a time resolution of 2
ns.

After switching off the trapping laser and the gradient magnetic field of the MOT, the atoms are
optically pumped to the initial state $|a\rangle$. The write pulse containing about $10^4$ photons
with a duration of 100 ns is applied into the atomic ensemble, to induce the spontaneous Raman
scattering via $|a\rangle\rightarrow|e\rangle\rightarrow|b\rangle$. The state of the induced
anti-Stokes field and the collective spin in Eq. (\ref{eqn:state}) is generated with a probability
$\chi\ll 1$. After a controllable delay of $\delta t$, the read pulse with the duration of 75 ns is
applied for converting the collective excitation into the Stokes field. In comparison, the
intensity of the read pulse is about 100 times stronger than that of the write one.

\begin{figure}[tb]
\includegraphics[width=4.2cm]{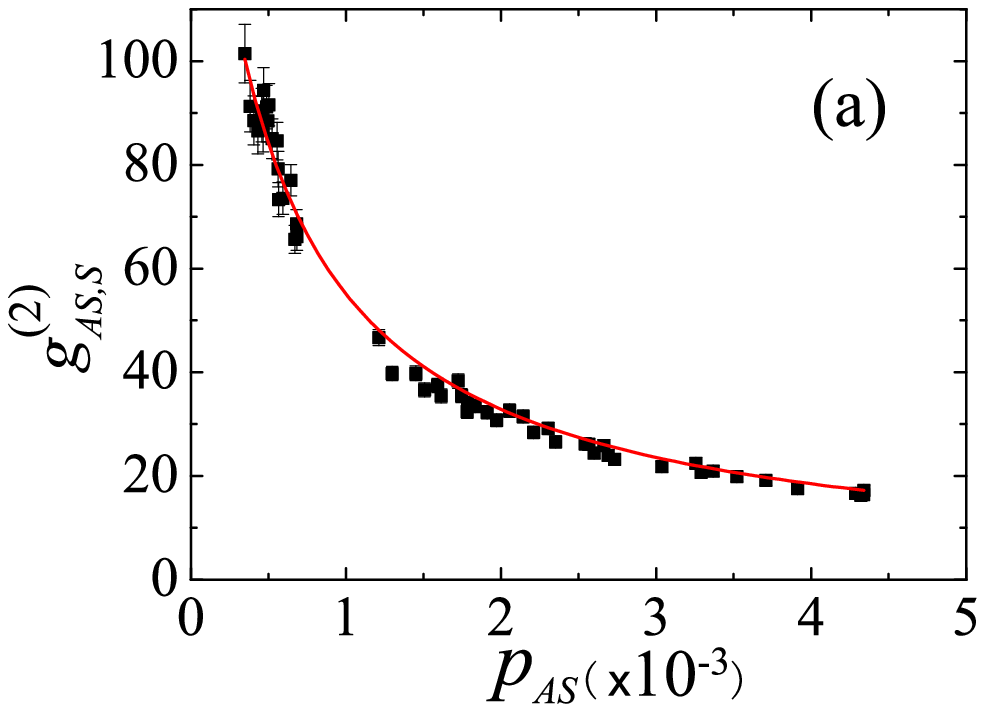}
\includegraphics[width=4.2cm]{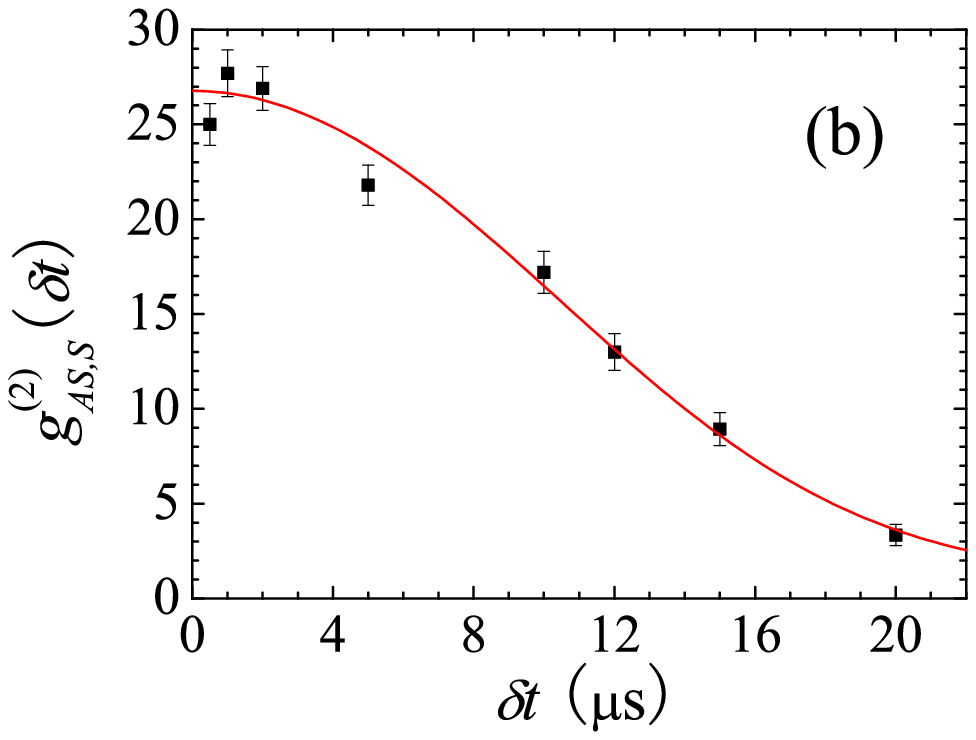}
\caption{Intensity correlation function $g_{\scriptscriptstyle AS,S}^{(2)}$ along the excitation
probability $p_{\scriptscriptstyle AS}$ with $\delta t=500$ ns (a) and along the time delay $\delta
t$ between read and write pulses with $p_{\scriptscriptstyle AS}= 3\times10^{-3}$(b). The black
dots are obtained from current experiment and the curves correspond to a least-square fitting
procedure according to Eq. (\ref{eqn:corr}) and (\ref{eqn:g12}). Here the observed lifetime is
$\tau_c=12.5\pm 2.6$ $\mu$s according to Eq.(\ref{eqn:g12}). \label{fig:lifetime}}
\end{figure}

Assume the probability to have an anti-Stokes (Stokes) photon is $p_{\scriptscriptstyle AS}$
($p_{\scriptscriptstyle S}$), and the coincident probability between the Stokes and anti-Stokes
channels is $p_{\scriptscriptstyle AS,S}$, then the intensity correlation function
$g_{\scriptscriptstyle AS,S}^{(2)}=p_{\scriptscriptstyle AS,S}/(p_{\scriptscriptstyle AS}
p_{\scriptscriptstyle S})=1+1/\chi$. We measured the variation of $g_{\scriptscriptstyle
AS,S}^{(2)}$ as a function of $p_{\scriptscriptstyle AS}$ shown in Fig. \ref{fig:lifetime}(a) with
a time delay of $\delta t=500$ ns. Considering the background in each channel, we obtain
\begin{subequations}\label{eqn:corr}
\begin{align}
p_{\scriptscriptstyle AS}& =\chi\eta_{\scriptscriptstyle AS}+B \eta_{\scriptscriptstyle AS},  \\
p_{\scriptscriptstyle S} & =\chi\gamma \eta_{\scriptscriptstyle S}+C \eta_{\scriptscriptstyle S}, \\
p_{\scriptscriptstyle AS,S} & =\chi \gamma \eta_{\scriptscriptstyle AS} \eta_{\scriptscriptstyle
S}+p_{\scriptscriptstyle AS} p_{\scriptscriptstyle S}.
\end{align}
\end{subequations}
Here, $\eta_{\scriptscriptstyle AS}$ and $\eta_{\scriptscriptstyle S}$ are the overall detection
efficiencies in the anti-Stokes and Stokes channels respectively, which include the transmission
efficiency $\eta_t$ of the filters and optical components, the coupling efficiency $\eta_c$ of the
fiber couplers, and the quantum efficiency $\eta_q$ of single photon detectors (there is another
spatial mode-match efficiency $\eta_m$ embodied in $\eta_{\scriptscriptstyle AS}$
\cite{ChaneliereNature2005}), $\gamma$ is the retrieve efficiency which is a time-dependent factor,
and $B$ ($C$) is a fitted factor indicating the background in the anti-Stokes (Stokes) channel.
The red curve is the least-square fitting result according to Eq. (\ref{eqn:corr}), where we assume
$B=0$ for simplicity. The efficiency in the anti-Stokes channel is observed as
$\eta_{\scriptscriptstyle AS}\sim 0.07$ and the retrieve efficiency $\gamma\sim 0.3$. It can be
seen the largest correlation $g_{\scriptscriptstyle AS,S}^{(2)}$ ($101\pm 6$) appears at the lowest
excitation probability $p_{\scriptscriptstyle AS}$ ($3.5\times 10^{-4}$).


The finite lifetime of the spin excitation results from from the dephasing of the collective state
due to the Larmor precession of the collective spin in the residual magnetic field. It can be
characterized by the decay of the retrieve efficiency $\gamma(\delta t)=\gamma_0 \exp(-\delta
t^2/\tau_c^2)$ \cite{ChaneliereNature2005}, where $\tau_c$ is the lifetime of the collective state.
Experimentally, the lifetime can be determined from the decay of the intensity correlation function
$g_{\scriptscriptstyle AS,S}^{(2)}(\delta t)$. Using Eq. (\ref{eqn:corr}), the intensity
correlation function reads
\begin{eqnarray}\label{eqn:g12}
g_{\scriptscriptstyle AS,S}^{(2)}(\delta t)=1+\frac{\gamma(\delta t)}{(B+\chi)\gamma(\delta t)+D},
\end{eqnarray}
where $C$ is absorbed by the new constant $D$. The decay of intensity correlation function
$g_{\scriptscriptstyle AS,S}^{(2)}(\delta t)$ is shown in Fig. \ref{fig:lifetime}(b). This curve
was taken at $p_{\scriptscriptstyle AS}=0.003$. The observed lifetime is $\tau_c=12.5\pm 2.6$
$\mu$s. It might be the noise arising from the elastic scattering of the write beam which makes the
cross correlation of the first point slightly lower than those of the two succeeding ones.


The protocol of the feedback process aims for a storable and time controllable single photon
source. As shown in Fig. \ref{fig:setup}(b), in a duration of $\Delta T$, $N$ independent write
pulses with a period of $\delta t_{\scriptscriptstyle W}$ are applied into the atomic ensemble.
Once an anti-Stokes photon is detected by D1 after one of the write pulses, the feedback circuit
stops further write pulses and enables the read pulse to retrieve the single photon after a time
delay $\Delta t$. If there is no click in D1, the atomic ensemble will be pumped back to the
initial state by a cleaning pulse. The above process will be halted until either an anti-Stokes
photon is detected, or the maximum number of trials ($N$) given by the lifetime of the excitation
is exceeded. In principle, the production rate of Stokes photons is enhanced while the
single-photon quality is conserved. This can be understood by the new excitation probability
$P_{\scriptscriptstyle AS}=\sum_{i=0}^{N-1}p_{\scriptscriptstyle AS}(1-p_{\scriptscriptstyle
AS})^i$.

Our protocol can be executed in two different modes. In the first mode, we fix the retrieving time
$\Delta T$. Therefore, the delay $\Delta t$ varies because the spin excitation is created randomly
in the $N$ write pulses. Single photons could be produced at a certain time with a high
probability, ideally approaching unity if $N\gg 1$. Furthermore, the retrieve efficiency can be
improved significantly by an increased optical depth of the atomic ensemble and an optimal retrieve
protocol \cite{AlexeyQuant0604037}. Thus, this mode serves as a deterministic single-photon source.
In second mode, we retrieve the single photon with a fixed delay $\Delta t$ after a successful
write. Thus the imprinted single photon can be retrieved at any time needed. This is well suited
for a quantum repeater \cite{BriegelPRL1998, DuanNature2001} when one node is prepared while
waiting for the other node.

In the first mode, we fixed $\Delta T=12.5$ $\mu$s and $\delta t_{\scriptscriptstyle W}=1$ $\mu$s.
Then $N=12$ write pulses were introduced. The anti-correlation parameter $\alpha$
\cite{GrangierEuLett1986} of field $\hat{a}_{\scriptscriptstyle S}$ characterizes the quality of
the single photon source. A 50/50 beam splitter is put in the Stokes channel to measure the
auto-correlation of the conditioned Stokes photons. The coincident probability between D1 and D2
(D3) is $p_{2,\scriptscriptstyle AS}$ ($p_{3,\scriptscriptstyle AS}$) and the three-fold coincident
probability in D1, D2 and D3 is $p_{23,\scriptscriptstyle AS}$ if we use only one write pulse and
retrieve immediately after the write. When we use $N$ write pulses and the feedback protocol, the
detection probabilities in D2 and D3 conditioned on a registration of an anti-Stokes photon in D1
are
\begin{eqnarray}\label{eqn:p2|1}
P_{m|{\scriptscriptstyle AS}}=\frac{\sum_{i=0}^{N-1}p_{\scriptscriptstyle
AS}(1-p_{\scriptscriptstyle AS})^i p_{m|{\scriptscriptstyle AS}}(\Delta T-n\cdot\delta
t_{\scriptscriptstyle W})}{\sum_{i=0}^{N-1}p_{\scriptscriptstyle AS}(1-p_{\scriptscriptstyle
AS})^i},
\end{eqnarray}
where $m=2,~3,~23$ and $p_{m|{\scriptscriptstyle AS}}(\Delta T-n\cdot\delta t_{\scriptscriptstyle
W})$ is a time-dependent probability conditioning on a click in the anti-Stokes channel. Therefore,
the anti-correlation parameter $\alpha={P_{23|{\scriptscriptstyle AS}}}/({P_{2|{\scriptscriptstyle
AS}}P_{3|{\scriptscriptstyle AS}}})$.

\begin{figure}[tb]
\includegraphics[width=4.0cm,height=3.2cm]{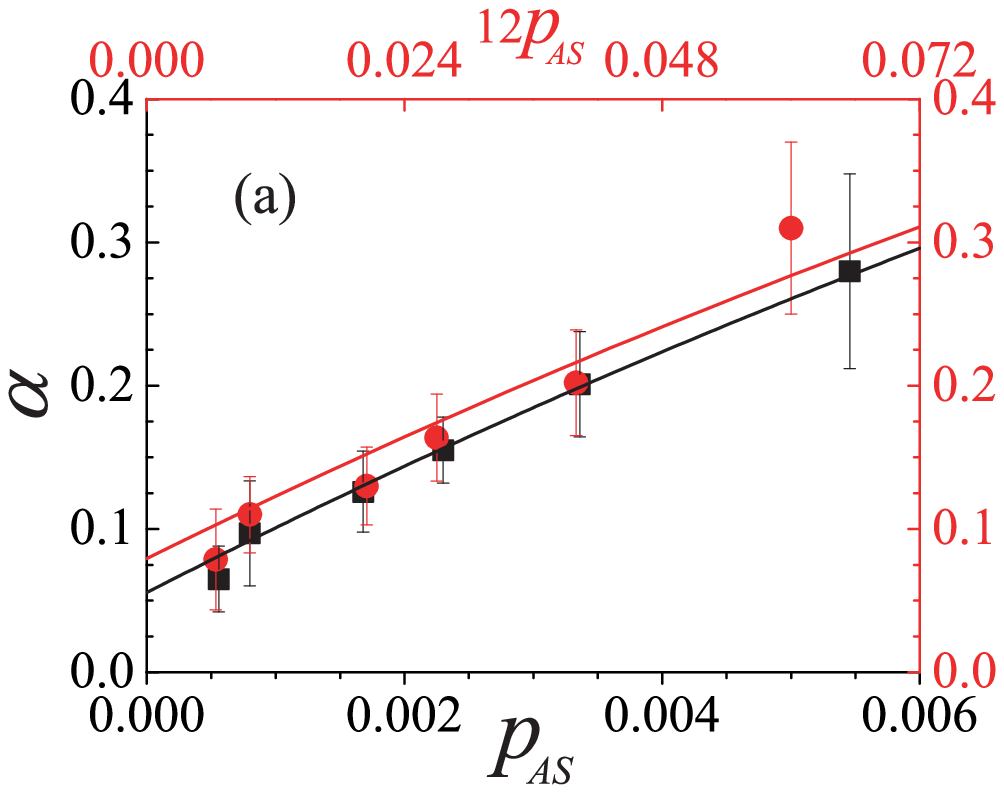}
\includegraphics[width=4.0cm,height=2.9cm]{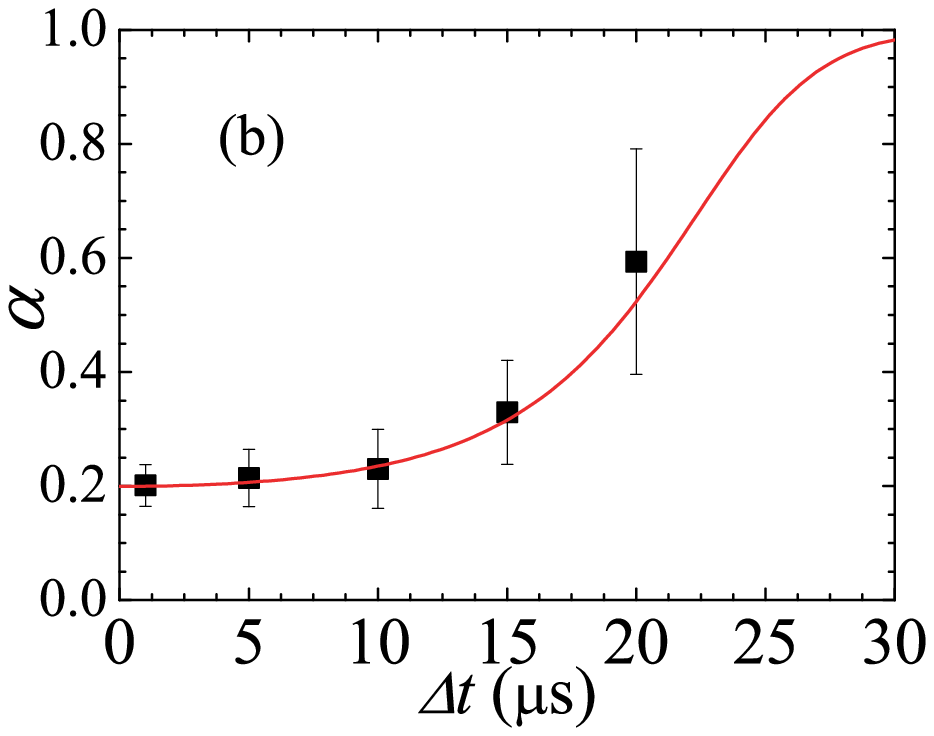}
\caption{The anti-correlation parameter as a function of $p_{\scriptscriptstyle AS}$ (a) and
$\Delta t$ (b). In Fig (a), the data in black correspond to the experiment without feedback
circuit, in which each write pulse is followed by one read pulse. Contrastively, the data in red
correspond to the experiment with feedback circuit, in which 12 successive write pulses are
followed by on read pulse. It can be seen that the single-photon nature is conserved even when we
increase the excitation probability $p_{\scriptscriptstyle AS}$ by 12 times. The red curve is the
theoretical evaluation taking account of the fitted background of the black dots, which are
consistent with the measured data. In Fig (b), 12 write pulses were applied in each trial while
measuring. When the delay $\Delta t < \tau_c$, the value of $\alpha$ keeps at a low level.
\label{fig:g22vsp1}}
\end{figure}

In Fig. \ref{fig:g22vsp1}(a), $\alpha$ was measured as a function of the excitation probability
$p_{\scriptscriptstyle AS}$. The variation of $\alpha$, shown as black dots, is nearly linear in
the region of $p_{\scriptscriptstyle AS}=0\sim 0.006$ when the experiment is performed as each
write pulse is followed by one read pulse. The black curve is the fitted result according to
Eq.(\ref{eqn:p2|1}) ($N=1$). When we use 12 successive write pulse, $\alpha$ versus
12$p_{\scriptscriptstyle AS}$ is plotted as red dots, which is consistent with the theoretical
curve evaluated from Eq. (\ref{eqn:p2|1}) ($N=12$) taking account of the fitted background in the
black curve. Note that, the value of $\alpha$ is $0.057\pm0.028$ when $p_{\scriptscriptstyle
AS}\rightarrow 0$, which should be 0 in principle. This offset comes from noise including residual
leakage of the write and read beams, the stray light, and dark counts of the detectors. However,
the advantage of the feedback protocol does not suffer from such noises. It is verified that
$\alpha$ is conserved even the excitation probability is much larger.
If the lifetime of the spin excitation is long enough to hold many write pulses, the excitation
probability can reach unity while the single-photon nature is still conserved. Then the generation
efficiency only depends on the retrieve efficiency itself.

In the second mode, $\delta t_{\scriptscriptstyle W}=1$ $\mu$s and $N=12$. In Fig.
\ref{fig:g22vsp1}(b), $\alpha$ was measured as the function of $\Delta t$. For every $\Delta t$,
$\Delta T$ varies due to the spin excitation created randomly in the $N$ write pulses. The behavior
$\alpha(\Delta t)$ is just like a reversed profile of $g_{\scriptscriptstyle AS,S}^{(2)}(\delta)$
in Fig. \ref{fig:lifetime}(b). However, when the delay $\Delta t < \tau_c$, the value of $\alpha$
keeps at a low level and varies more slowly compared with $g_{\scriptscriptstyle AS,S}^{(2)}$. Even
when we extend the delay to 20 $\mu$s, $\alpha\sim 0.6$ which is still smaller than 1. Satisfying
agreement is observed between the theoretical curve and the experimental data.

As demonstrated in the present work, the lifetime of collective states is important for the quality
and production rate of single photons. In the atomic ensemble, the coherence time of the collective
state suffers from the residual magnetic field around the MOT and the thermal motion of the atoms.
The latter effect is negligible because of the very low temperature of the atomic cloud. Using a
better compensation of residual magnetic field (within current technology, below 1mG) we can
greatly increase the lifetime of the collective state. Moreover, by further improving the control
circuit, i.e. reducing the period of write pulses, we can apply more write pulses within the
lifetime. In particular, in the case with $p_{\scriptscriptstyle AS}=0.003$ and a write period of
300 ns, we can obtain a single-photon source with a probability as high as 95\% within a lifetime
of 300 $\mu$s.

In conclusion, we have demonstrated an experimental realization of an controllable single-photon
source with atomic storage. The lifetime of the collective spin excitation reaches 12.5 $\mu$s. A
feedback circuit was constructed to control the generation of the spin excitation and the storage
time $\delta t$. Being a key device in the scalable quantum communication network, this circuit
also shows a promising performance in the enhancement of the excitation probability while the
single-photon quality is conserved. This single-photon source is able to work at either a
deterministic mode or a time controllable mode heralded by the feedback circuit. The single-photon
source based on atomic ensemble has the advantages of narrow band, high quality and controllable
character, which is helpful for the construction of scalable quantum information processing system
in the future.

This work was supported by the Deutsche Forschungsgemeinschaft (DFG), the Alexander von Humboldt
Foundation, the Deutsche Telekom Stiftung and the Konrad-Adenauer Stiftung.

\textit{Note added}.-- During the preparation of our manuscript, we are aware of two recently
related experiments by A. Kuzmich's group \cite{MatsukevichQuant0605098} and H. J. Kimble's group
\cite{LauratQuant0605122}.

\end{document}